\documentclass[%
 reprint,
 amsmath,amssymb,
 aps,
]{revtex4-1}

\usepackage{graphicx}
\usepackage{dcolumn}
\usepackage{bm}
\usepackage{hyperref}
\usepackage{aas_macros}

\begin{document}

\preprint{APS/123-QED}

\title{Density Fluctuation Reconstruction using KS test and D'Agostino's K-squared test}
\thanks{E-mail: yoshida.kiichi@f.mbox.nagoya-u.ac.jp}%

\author{Kiichi Yoshida}
 \affiliation{Department of Physics, Nagoya University, Furocho Chikusaku Nagoya, 464-8602 Aichi, Japan}
\author{Kiyotomo Ichiki}%
\affiliation{Department of Physics, Nagoya University, Furocho Chikusaku Nagoya, 464-8602 Aichi, Japan\\Kobayashi Maskawa Institute, Nagoya University, Furocho Chikusaku Nagoya, 464-8602 Aichi, Japan}
\author{Atsushi J. Nishizawa}
\affiliation{Institute for Advanced Research, Nagoya University, Furocho Chikusaku Nagoya, 464-8602 Aichi, Japan}




\date{\today}

\begin{abstract}
We propose a new non-parametric method to constrain the cosmological model
through the growth factor of large-scale structure. 
To constrain the cosmological model from observations such as cosmic microwave background or large-scale structure, the power spectrum or correlation function have been used since they are well described by the linear perturbation theory or little correction is needed for the non-linearity; however, it is always important to find an alternative route to constrain cosmological models in order to avoid systematic effects due to the estimator and make the constraint more robust.
Our density fluctuation reconstruction method constrains the cosmological model directly from time evolution of density fluctuation.
We propose to apply Kolmogorov-Smirnov test and D'Agostino's K-squared test  to the distribution function of density fluctuations, which are observed on a light cone. The density fluctuation recovers the Gaussian distribution only if we reconstruct the density fluctuation with correct linear growth factors. The method is verified by the use of Gaussian random field and N-body simulations. This method will open a new window both to separate the initial three dimensional density fluctuation and the growth factor of large scale structure.
\end{abstract}

\maketitle

\section{Introduction}
\label{sec:introduction}

The Big-Bang cosmological model with cold dark matter and cosmological constant, so called the $\Lambda$CDM model, can explain various observational phenomena, such as abundances of the light elements produced by the Big-Bang Nucleosynthesis, the existence of the Cosmic Microwave Background (CMB), the dimming of distant Type Ia supernovae, and so on \citep{1998AJ....116.1009R,1999ApJ...517..565P,2013PhR...530...87W,1992ApJ...396L...1S,2003ApJS..148..175S,2018arXiv180706205P}.  In this standard cosmological model, the universe contains small density fluctuations, which are the seeds for the anisotropies of the CMB and the large scale structure of the universe observed today. Time evolution of the density fluctuations is described by the well-established cosmological perturbation theory, and by comparing theoretical predictions with observational data the cosmological parameters have been determined within a few percent level \citep{2018arXiv180706205P}.

Most of cosmological observations of density fluctuations denoted by $\delta$ are limited on our light cone and we are only able to obtain the information of $\delta(x,t)$ in "different place $x$ and different time $t$ [$\delta_{\rm obs}(x, t)$]" (see, e.g., \cite{1997ApJ...491L...1M,2000MNRAS.314...92T}). In the context of cosmological linear perturbation theory, the time evolution of density fluctuations can be described as $\delta(x,t) = D(t) \delta(x,t_{\rm ini})$ where $D(t)$ is the linear growth factor \citep{10.1046/j.1365-8711.2001.04137.x}. The growth factor depends on cosmological parameters and contains information about the universe. When we compare theoretical models to observations, the most conventional way to constrain cosmological parameters has been based on the two point statistics, namely, the power spectrum or two-point correlation functions of the density field since we are more interested in the growth of density perturbations $D(t)$ than their phase information in $\delta(x,t_{\rm ini})$. From the ratio of the variance of $\delta$ at two different epochs with an assumption that the variance of $\delta(x,t_{\rm ini})$ is common all over the space, we may estimate $D(t_2)/D(t_1)$. For example, by comparing the variances between in the CMB epoch and present, we may infer the existence of cold dark matter and be able to constrain its abundance $\Omega_{\rm m}$. 

Another popular method to investigate the evolution of the growth factor is based on observations of the peculiar velocity field \citep{2011MNRAS.417.3101P} since the divergence of the velocity field directly connects to the time evolution of the density field through the continuity equation.  Recent measurements of the velocity field from redshift space distortions such as eBOSS \citep{2018MNRAS.477.1604G, 2015MNRAS.449..848H, 2017MNRAS.470.2617A, 2017A&A...608A.130D, 2017A&A...603A..12B} have provided us with constraints on the growth rate, which are consistent with the Planck cosmology and general relativity. The velocity field amplitude is estimated from the matter power spectrum of monopole, quadrupole and hexadecapole since the peculiar velocity along the line of sight will make the galaxy power spectrum anisotropic.

In this paper, we develop a new method to estimate the time evolution of density fluctuations $D(t)$, without utilizing variance estimators. In order to separate $D(t)$ out from $\delta(x,t)$, we base our method on the working hypothesis that $\delta(x,t_{\rm ini})$ obeys Gaussian distribution but $\delta(x,t)$ does not. The Gaussianity of the initial density fluctuations (e.g. \cite{2010MNRAS.409..737W}) is a reliable assumption because primordial non-Gaussianity $f_{\rm NL}$ has been constrained in a very small value according to the recent Planck results \citep{2018arXiv180706205P}.  To estimate how the distribution $f(x)$ is different from proposed distribution $g(x)$, we use the Kolmogorov-Smirnov test (KS test, \cite{10030673552, Smirnov:1939:EDB}). And also we apply the D'Agostino's K-squared test \citep{10.2307/2334794} which diagnoses whether the sample is drawn from Gaussian distribution or not. 
Our method is non-parametric way \citep{2016JCAP...04..016G}, and therefore we can estimate the linear growth factor $D(t)$ without assuming any cosmological model. 

The structure of this paper is as follows. In section \ref{sec:Data}, we explain how we prepare the simulation data, using Random Gaussian simulations \citep{2005astro.ph..6540M}. We also describe N-body simulations \citep{article} to study the effect of non-linear evolution of density fluctuations. In section \ref{sec:Reconstruction Method}, we describe our reconstruction methods in two different prescriptions and we show the result in section \ref{sec:Result}. In section \ref{sec:Discussion}, we discuss the interpretation of our results and how precisely we may recover the growth factor using our method, and we conclude this work in section \ref{sec:Conclusion}. Throughout the paper, we assume the cosmological parameters consistent with the Planck 2018 results \citep{2018arXiv180706205P}.

\section{Mock data}
\label{sec:Data}
In this section, we describe the method to make a "observed" matter density field. We use 
two different data set, Random Gaussian \citep{2005astro.ph..6540M} and N-body simulation \citep{article}.


\subsection{Random Gaussian Simulation}
\label{ssec:Random Gaussian Simulation}

In order to verify our method, we first consider the simplest data set which is 
a random Gaussian density field.
First we consider to generate the matter density fluctuation $\delta(\bm{k})$ 
in Fourier space. In linear regime, initial density fluctuation obeys
Gaussian distribution. Therefore, we generate the density fluctuation at present $\delta(\bm{k}, z_0)$ such that 
it satisfies
%
\begin{equation}
    {\mathcal P}[\delta(\bm{k}, z_0)]
    =
    \mathcal{N}
    [\delta|\mu=0, \sigma^2=P(|\bm{k}|, z_0)],
    \label{ditribution of deltak}
\end{equation}
where $\mu$ and $\sigma^2$ are mean and variance of Gaussian probability distribution, respectively. $P(k, z_0)$ is a power spectrum of matter density at present time $z_0=0$ 
and we compute the $\Lambda$CDM power spectrum using publicly available code \texttt{CLASS}
\cite{2011arXiv1104.2932L,Blas_2011}. 
%
Then we Fourier transform to obtain three dimensional density fluctuation in a real space,
\begin{equation}
    \delta(\bm{x}, z_0)
    =
    \int_{k_{\rm min}}^{k_{\rm max}} \frac{d^3 \bm{k}}{(2\pi)^3}
    \delta(\bm{k}, z_0)e^{-i\bm{k}\bm{x}}. 
    \label{Fourier transform}
\end{equation}
We cut the Fourier transform at finite scales from $k_{\rm min}$ to $k_{\rm max}$ due to the practical computation, and therefore $\delta(\bm{x}, z_0)$ does not exactly follow the Gaussian distribution. However, as we will show later, we can approximate $\delta(\bm{x})$ as Gaussian variables 
because the contributions outside the range of interest are negligibly small. 
Now we consider the observed density fluctuation. In the linear regime, the observed density fluctuation can be written
\begin{equation}
    \delta_{\rm obs}(\bm{x}, z)=\delta(\bm{x}, z_0)
    D(z),
    \label{make observational data}
\end{equation}
where $D$ is a linear growth factor \citep[e.g.][]{1980lssu.book.....P}.  Since what we observe in practice is along the light path, we set the data vector as $\bm{X}=\{\delta(\bm{x}_1,z_1),\cdots \delta(\bm{x}_N,z_N) \}$. Then $\bm{X}$ no longer obeys Gaussian distribution, because $\delta$ at different redshift is sampled from Gaussian distribution which has different variances by the factor of $D^2$.
In order to generate a simulated data which encompasses this light cone effect,
we assume the standard $\Lambda$CDM cosmology to compute $D=D_{\rm true}(z)$, with $\Omega_m=0.3$ and $\Omega_K=0$.

We first define a cubic, 7 Gpc/$h$ on a side and put the observer at the center of the cubic. The density fluctuations at present $z=z_0$ are computed by equation \eqref{Fourier transform} at the position of grids which divide the volume into $100^3$ segments. Then we multiply the linear growth factor $D(z)$, corresponding to the comoving distance from the observer to each grid point.


\subsection{N-body simulation}
\label{ssec:N-body}

In order to consider the more realistic situation, we use dark matter N-body simulations.
Although the linear growth of density fields still holds on large scales, the non-linear gravitational evolution may exert a significant impact on the distribution function of $\delta$. Since our method is sensitive to the intrinsic distribution function of $\delta$, it is important to assess the validity of our method with the realistic density fields which incorporate the non-linear growth of structure.

We perform N-body simulation using publicly available \texttt{Gadget-2} code \citep{2005MNRAS.364.1105S}. The simulation box is $L_{\rm box}=2.2$ Gpc/$h$ and number of particles is $N_p=512^3$ such that we can trace the density field down to $22$ Mpc/$h$. The initial condition is created by the \texttt{2LPT} code \citep{2012ascl.soft01005C}, at $z=20$. 

We assume that the observer is located on a vertex of the simulated box and observes one-eighth 
of the whole sky.
We divide the simulation box into 96 spherical shell such that all shells have equal volume, $\Delta V = 4\pi \chi^2 \Delta \chi=4\pi\chi_0^3/3$, where $\chi_0$ is the comoving distance to the first shell's boundary.
As in the random Gaussian simulation, we divide the box into $100^3$ cells on which we compute the density fluctuation. The redshift of each shell is represented by the one corresponding to the comoving distance to the middle of the shell calculated by $(\chi_i+\chi_{i+1})/2$.

\begin{figure*}
    \centering
    \includegraphics[scale=0.6]{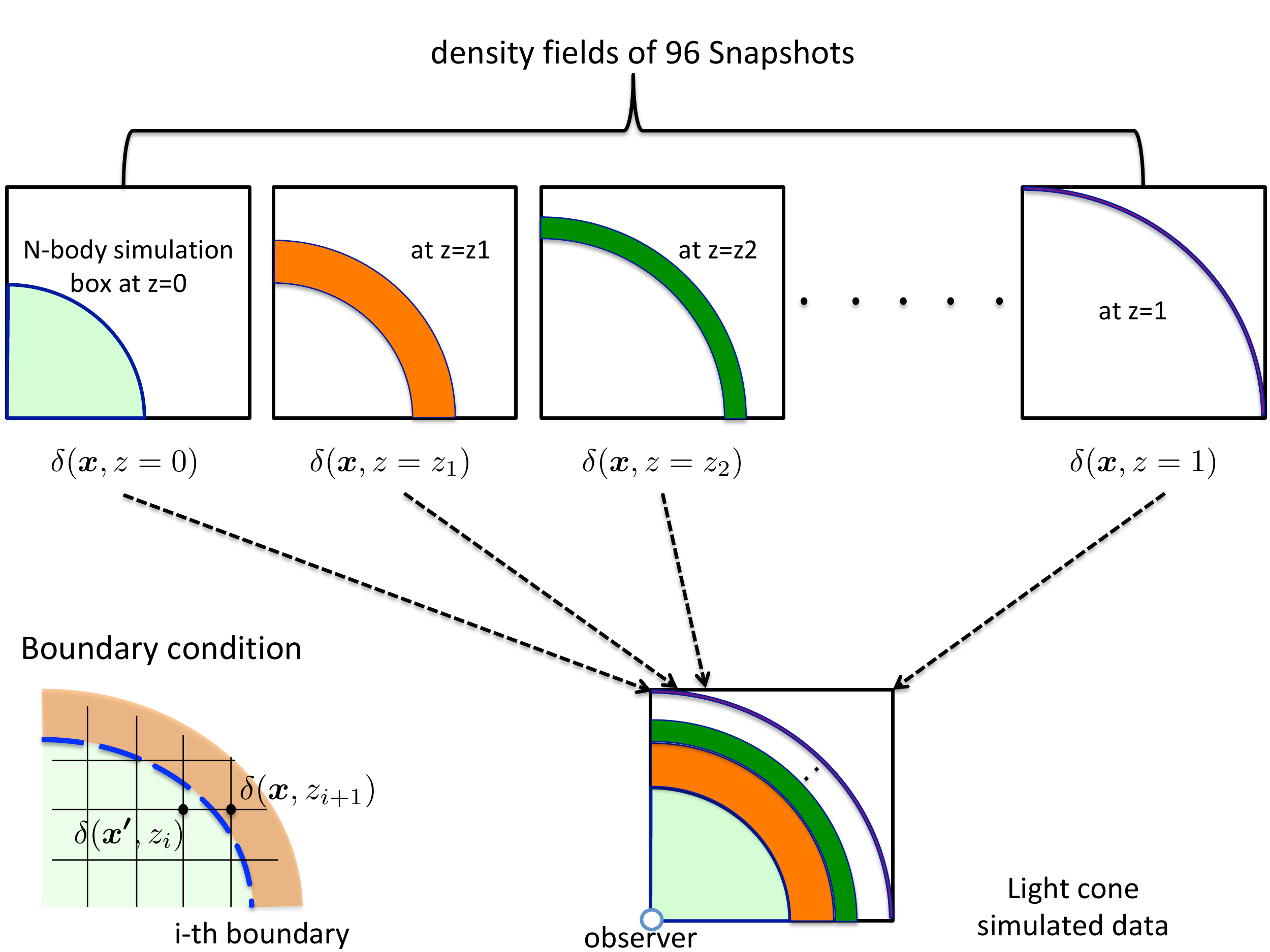}
    \caption{Graphical explanation how we make light-cone data $\delta_{\rm obs}(\bm{x}, z)$ from N-body simulation and how we treat boundary condition. We first make density fluctuation $\delta(\bm{x}, z_i)$ for each grid using $i$-th snapshot data and then divide the shell. The detail is discussed in section \ref{ssec:N-body}.}
    \label{fig:illustration_Nbody}
\end{figure*}

This means that we need to have snapshots at 96 different redshifts. The density fluctuation on the cell sitting within the $i$-th shell can be directly computed from the 
snapshot at z=$z_i$. We show schematic illustration how we generate the light-cone simulated data in Figure \ref{fig:illustration_Nbody}.

\section{Reconstruction Method}
\label{sec:Reconstruction Method}

Now we prepare simulated data $\delta_{\rm obs}$ in two different ways. The question here is how we can reconstruct $\delta(\bm{x}, z_0)$ and growth factor $D(z)$ simultaneously. It is in general impossible to solve this question because the number of variants to be obtained is less than the number of observables. In this work, we assume that the density fluctuation $\delta$ obeys Gaussian distribution to regularise the problem.

\subsection{Kolmogolov-Smirnov Test}
\label{ssec:KS test}
Kolmogorov-Smirnov test (hereafter KS test)
is 
one of the most broadly used method
to test if the two finite samples are drawn from same probability distribution
or to compare the distribution function of 
the sample
to the proposed distribution 
\citep{10030673552, Smirnov:1939:EDB}. 
Here we use the KS test to see if the distribution function of $\delta$ is inconsistent with Gaussian distribution. In this section, we briefly revisit the formalism of the KS test.

Suppose that we 
have $n$ 
data in hand
which is drawn from unknown 
distribution function $f(\delta)$ and 
that the data is arranged in an ascending order, as $\bm{\delta}=\{\delta_1, \delta_2, \cdots, \delta_n\}$, where $\delta_i<\delta_{i+1}$ for any integer $i$. The cumulative distribution function of this sample can be formally expressed as
\begin{equation}
    F_n(\delta)=\frac{1}{n}\sum^n_i \delta_i\Theta(\delta-\delta_i),
    \label{EDF}
\end{equation}
where $\Theta(y)$ is a Heaviside step function. 
 Now our concern is if the unknown distribution function measured from the data is consistent with the proposed distribution function, $g(\delta)$. The cumulative distribution function of $g(\delta)$ is defined as
\begin{equation}
    G(\delta)=\int^{\delta}_{-\infty} g(\delta') d\delta'.
    \label{cumulative function}
\end{equation}
The null hypothesis to be tested is that "$F_n(\delta)$ is the same distribution as $G(\delta)$". In order to see this, the statistical value $S_n$ is introduced,
\begin{equation}
    S_n=\sup_\delta |F_n(\delta)-G(\delta)|,
    \label{statistical value}
\end{equation}
where $\sup$ denotes the supremum.
The $p$-value which will be used for test the hypothesis can be given as
\begin{equation}
    p(\sqrt{n}S_n)=2\sum^{\infty}_{m=1}(-1)^{m-1} e^{-2m^2(\sqrt{n}S_n)^2},
    \label{p_value}
\end{equation}
which is called "Kolmogorov distribution" \citep{10030673552}. If we set significance level $\alpha\%$ and define $\beta$ which satisfies $p(\beta)=\alpha/100$, then the null hypothesis will be rejected when $\sqrt{n}S_n>\beta$. $p$-value is the cumulative expression of the likelihood  function $\mathcal{L}$, so we simply take derivative of equation \eqref{p_value}.
\begin{equation}
    \mathcal{L}(\sqrt{n}S_n)=-\left.\frac{dp}{dx}\right|_{x=\sqrt{n}S_n}.
    \label{Likelihood}
\end{equation}

In the following section, we will use the likelihood function $\mathcal{L}$ to quantify the degree of disagreement from the true underlying probability distribution.

\subsection{D'Agostino's K-squared Test}
\label{ssec:DA test}
D'Agostino's K-squared test can be used for 
discriminate whether the sample is drawn from
Gaussian distribution or not, based on the higher moments of the samples.\par
The skewness $g_1$ and kurtosis $g_2$ for the sample can be defined as
\begin{align}
    \label{e:skewness}
    &g_1=
    \frac{\langle (\delta_i-\bar{\delta})^3 \rangle}{\langle (\delta_i-\bar{\delta})^2 \rangle ^{3/2}},\\
    \label{e:kurtosis}
    &g_2=
    \frac{\langle (\delta_i-\bar{\delta})^4 \rangle}{\langle (\delta_i-\bar{\delta})^2\rangle^2}-3,
\end{align}
where $\langle \cdots \rangle$ stands for the arithmetic mean over $n$ samples. According to \cite{10.2307/2332104}, if the sample is drawn from Gaussian distribution, then the distribution of $g_1$ should have the mean $\mu$, variance $\sigma^2$, skewness $\gamma_1$ and kurtosis $\gamma_2$ as
\begin{align}
    &\mu(g_1)=0,\\
    &\sigma(g_1)=\frac{6(n-2)}{(n+1)(n+3)},\\
    &\gamma_1(g_1)=0,\\
    &\gamma_2(g_1)=\frac{36(n-7)(n^2+2n-5)}{(n-2)(n+5)(n+7)(n+9)},
\end{align}
and similarly, for $g_2$ we have
\begin{widetext}
\begin{align}
    &\mu(g_2)=-\frac{6}{n+1},\\
    &\sigma(g_2)=\frac{24n(n-2)(n-3)}{(n+1)^2(n+3)(n+5)},\\
    &\gamma_1(g_2)=\frac{6(n^2-5n+2)}{(n+7)(n+9)}\sqrt{\frac{6(n+3)(n+5)}{n(n-2)(n-3)}},\\
    &\gamma_2(g_2)=\frac{36(15n^6-36n^5-628n^4+982n^3+5777n^2-6402n+900)}{n(n-3)(n-2)(n+7)(n+9)(n+11)(n+13)}.
\end{align}
\end{widetext}
The distributions of $g_1$ and $g_2$ become close to 
the Gaussian as the sample number $n$ increases. 
To treat them more precisely, \cite{10.2307/2334794} has transformed $g_1$ to the so called $Z$-value 
defined as
\begin{equation}
    Z_1(g_1)=\Delta{\rm ln}\left(\frac{g_1}{\alpha\sqrt{\sigma(g_1)}}+\sqrt{\frac{g^2_1}{\alpha^2\sigma(g_1)}+1}\right),
    \label{e:z_1 value}
\end{equation}
where $\Delta=1/\sqrt{{\rm ln} W}, \alpha^2=2/(W^2-1)$ and $W^2=\sqrt{2\gamma_2(g_1)+4}-1$. And \cite{10.2307/2684359} has also proposed the transformation 
of $g_2$ to $Z$-value as
\begin{widetext}
\begin{equation}
    Z_2(g_2)=\sqrt{\frac{9A}{2}}\left[1-\frac{2}{9A}-\left(\frac{1-2/A}{1+\frac{g_2-\mu(g_2)}{\sqrt{\sigma(g_2)}}\sqrt{2/(A-4)}}\right)^{1/3}\right],
    \label{e:Z_2 value}
\end{equation}
\end{widetext}
where
\begin{equation}
    A=6+\frac{8}{\gamma_1(g_2)}\left(\frac{2}{\gamma(g_2)}+\sqrt{1+4/\gamma^2_1(g_2)}\right).
    \label{e:factor}
\end{equation}
From these $Z$-values, we can get test statistic
as
\begin{equation}
    K^2=Z_1^2(g_1)+Z_2^2(g_2).
    \label{e:K-value}
\end{equation}
Because the $Z$-value obeys Gaussian distribution, $K^2$ obeys chi-square distribution defined as
\begin{equation}
    P(K^2, \nu)=
    2^{-\nu/2}\Gamma^{-1}\left(\frac{\nu}{2}\right)
    K^{\nu-2}{\rm exp}\left(-\frac{K^2}{2}\right),
    \label{e:chi square distribution}
\end{equation}
where $\nu$ is the degree of freedom and $\Gamma(x)$ is the Gamma function and in this case, $\nu=2$. 
Thus, we can obtain the $p$-value or likelihood $\mathcal{L}$ using equation \eqref{e:K-value}, \eqref{e:chi square distribution} and calculate the probability 
that the sample obeys the Gaussian distribution.

\subsection{Apply for Random Gaussian Simulation}
\label{ssec:Apply for Random Gaussian Simulation}
\begin{figure}
 \centering
 \includegraphics[width=8cm,clip]{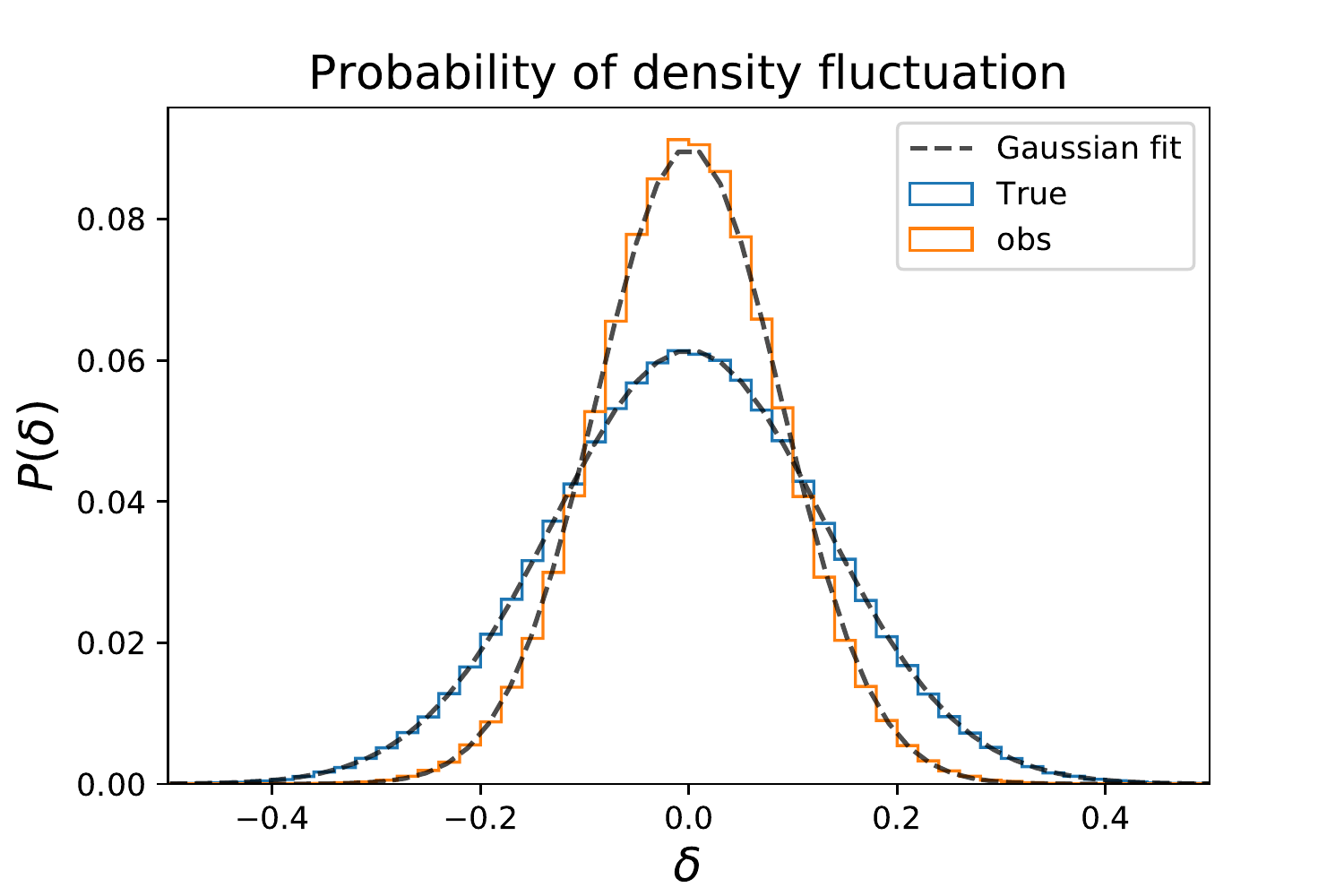}
 \caption{Probability distribution of density fluctuation from Random Gaussian simulation. Blue histogram shows the probability distribution of $\delta(\bm{x}, z_0)$ and orange one shows  probability distribution of $\delta_{\rm obs}(\bm{x}, z)$. Dashed line show the Gaussian-fitted probability distribution function (PDF). We used density fluctuation only within the radius $3.5$ Gpc/$h$ in both cases. }
 
 \label{HIst_comp_randomG}
\end{figure}

In this section, we demonstrate that how the KS test and D'Agostino's K-squared test work for reconstruction of the density field by the simplest data set of the random Gaussian simulation.
Figure \ref{HIst_comp_randomG} shows the probability distribution of $\delta(\bm{x}, z_0)$ at present time and $\delta_{\rm obs}(\bm{x}, z)$. 
Since $\delta(\bm{x}, z_0)$ is a random draw from the single Gaussian distribution, the $p$-value of the KS test and D'Agostino's test are $0.93$, and $0.381$, which are far from rejecting the Gaussian distribution in both cases. Conversely, since $\delta_{\rm obs}(\bm{x}, z)$ contains density fluctuation at different epochs and thus taken from multiple Gaussian distributions where the variances are different, the $p$-values of the KS test and D'Agostino's test are $2.76\times 10^{-7}$ and $1\times 10^{-117}$, which fairly rejects the Gaussian hypothesis.

To reconstruct density fluctuation 
and growth factor, we 
divide the redshift range into 9 bins such that each shell has the equal comoving volume and leave the growth rate at each shell free parameter without assuming any cosmological models; i.e. $\bm{D}=\{D_0, D_1, \cdots, D_8\}$. The growth factor parameters are estimated by the Markov Chain Monte Carlo method (MCMC) \citep[e.g.][]{MacKay:2002:ITI:971143} with the Metropolis-Hastings  (MH) algorithm. We fix $D_0=1$ and simultaneously estimate 8 other parameters since our method is insensitive to the overall amplitude.

In the MH method, the proposed parameters are accepted according to the likelihood ratio, and we apply the likelihood function calculated from equation \eqref{Likelihood}. We accept the proposed parameter set if 
\begin{equation}
    \alpha = \frac{\mathcal{L}({\rm proposed})}{\mathcal{L}({\rm current})} > r,
\end{equation}

where $r$ is a random number ranging from 0 to 1, which is drawn from the uniform distribution in every step of the MCMC. 
For our D'Agostino's test, we simply use equation \eqref{e:chi square distribution} for the likelihood function.


\subsection{Apply for N-body simulation}
\label{ssec:Apply for N-body simulation}
%
\begin{figure}
 \centering
 \includegraphics[width=8cm,clip]{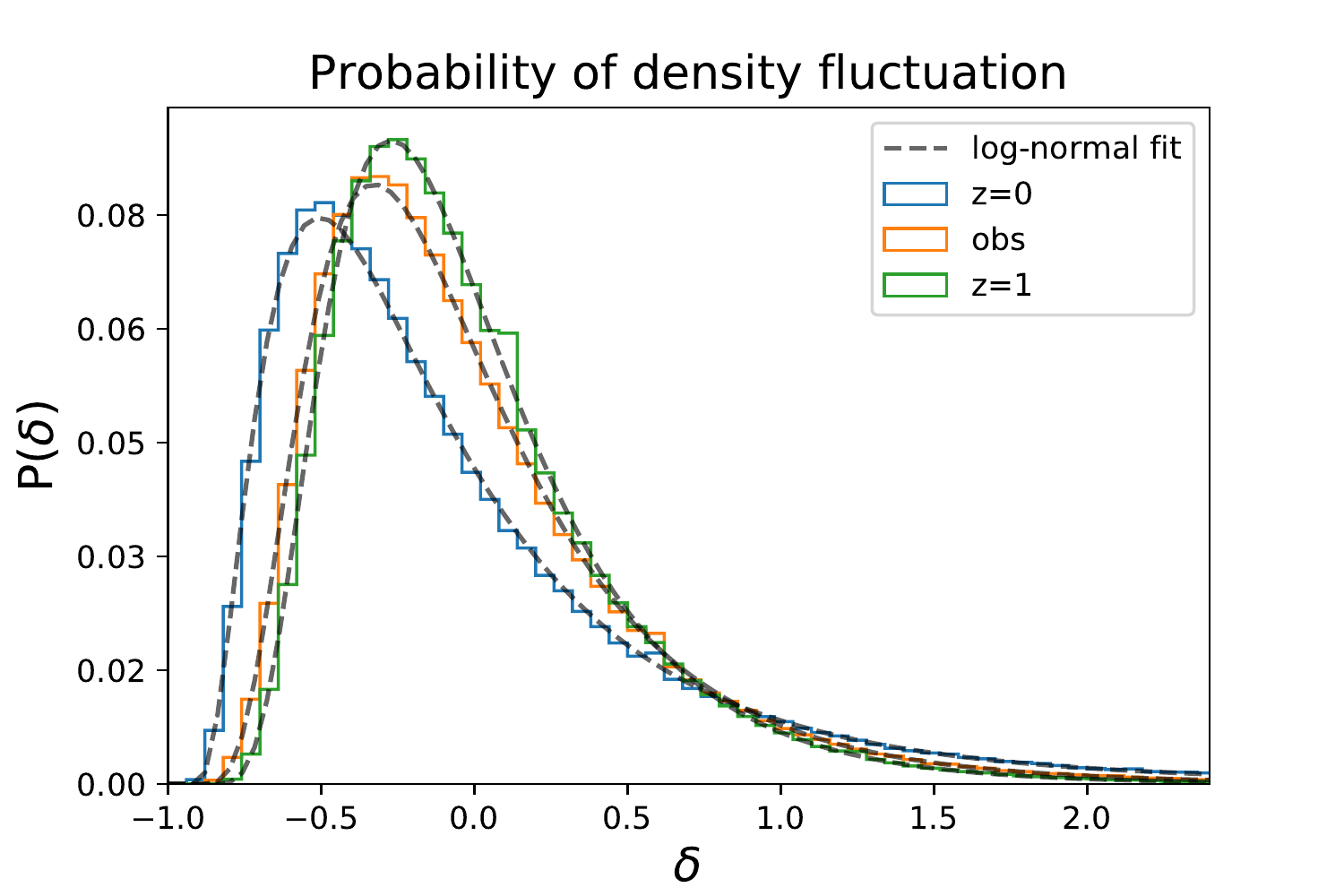}
 \caption{Probability of density fluctuation from N-body simulation. Blue histogram shows the probability distribution of $\delta(\bm{x}, z_0)$ and orange one shows $\delta_{\rm obs}(\bm{x}, z)$ and green one is $\delta(\bm{x}, z_{\rm max}=1)$. Dashed lines show the log-normal fitted PDF. We used data points only within $z_{\rm max}$}.
 
 \label{HIst_comp_N}
\end{figure}
%
For more realistic situation, we consider the density fluctuations generated using N-body simulations.
Figure \ref{HIst_comp_N} shows the histograms for density fluctuations from our N-body simulations. 
It is well known that the density fluctuation of large-scale structure is well described by the log-normal distribution (e.g. \cite{Kayo_2001}). Since the distribution is highly skewed and not close to the Gaussian, we cannot directly apply the KS test or D'Agostino's test to this sample.



In order to apply our methods to this skewed data, we propose two different approaches. 
The first approach is that we directly apply the KS test to the data. As we mentioned above, the density fluctuation at each redshift is well described by the log-normal distribution. Although we find that the $p$-value for $\delta(\bm{x}, z=1)$ 
to the best-fitting log-normal distribution is $3.6\times 10^{-37}$ and conclude that the distribution is far from log-normal distribution, what we need to test is the relative significance of rejecting the distribution. In other words, even if the underlying distribution of $\delta$ is far from the log-normal distribution, as long as the wrong combination of $\bm{D}$'s gets smaller $p$-value, it means that the KS test still has ability to constrain the parameters.
We assign parameter $\bm{D}$ in 10 redshift bins and fix $D_9(z_9)=1$.

\begin{figure}
 \centering
 \includegraphics[width=8cm,clip]{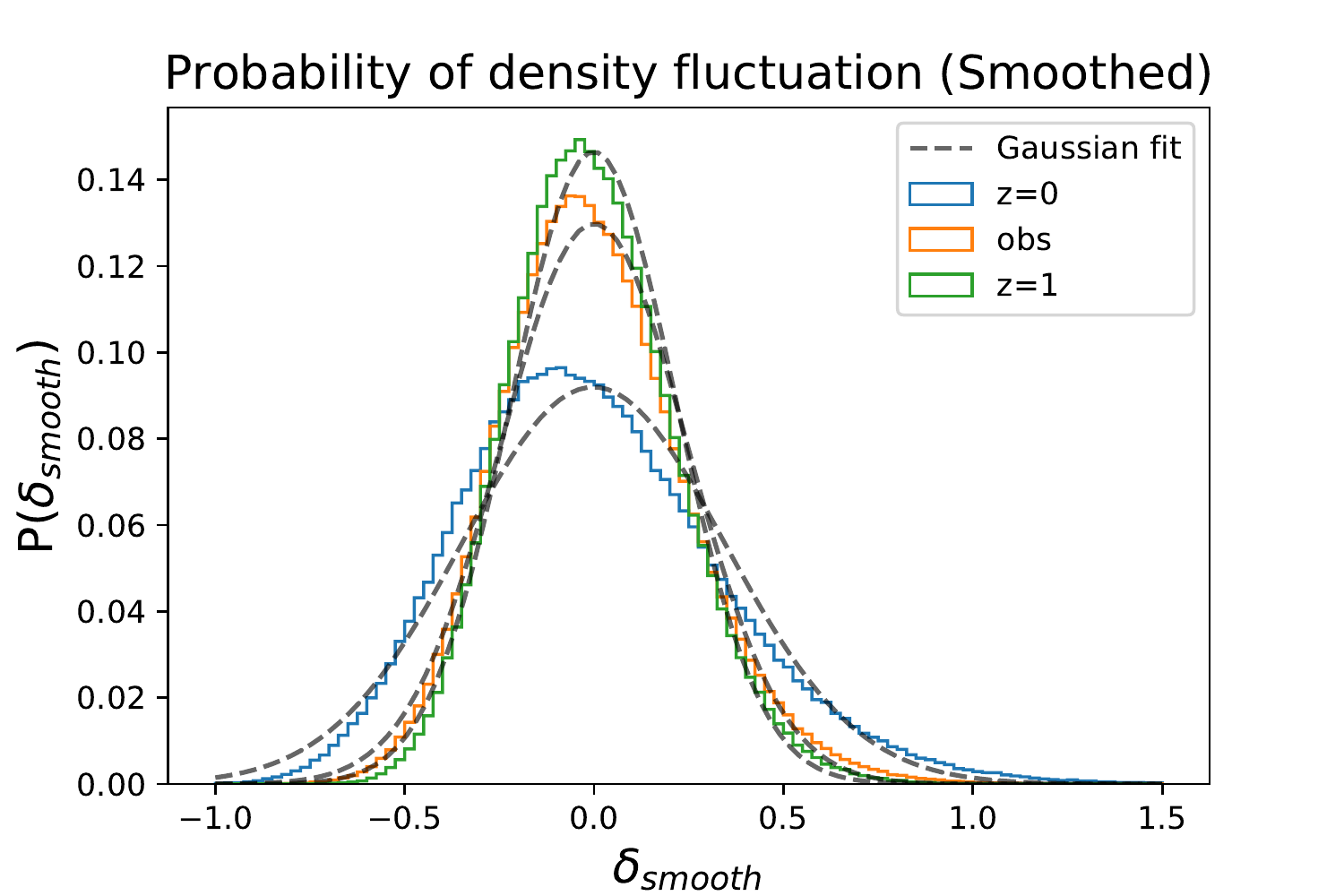}
 \caption{
 Same as Figure \ref{HIst_comp_N} but for the smoothed fluctuation with $k_{\rm cut}=0.07\ h/{\rm Mpc}$. 
 }
 \label{HIst_comp_N_sm}
\end{figure}

The second approach is that we only use the large-scale mode fluctuations to make the distributions closer to the Gaussian distribution.
We apply a top-hat truncation in Fourier space as $\delta_{\rm sm}(\bm{k})=\delta(\bm{k})\Theta(k_{\rm cut}-k)$, where we adopt the value $k_{\rm cut}=0.07\ h/{\rm Mpc}$. 
Figure \ref{HIst_comp_N_sm} shows the distribution after applying the smoothing. We assign parameter $\bm{D}$ in 10 redshift bins as in the case of the first approach. However, at the lowest redshift bin, the non-linear effect is still visible even after smoothing, and thus we discard the lowest redshift bin and fix the growth factor as $D_9(z_9)=1$. \par
In table.\ref{table:redshift}, we summarize the redshift range and the volume of these simulations.

\begin{table*}
\begin{center}
\caption{Radshift range and the volume of each parameters. In N-body simulation pipeline 2, we dont't use the first redshift bin $z_0$.}
\begin{tabular}{c|cc|cc}\hline
 & \multicolumn{2}{c|}{Random Gaussian (full sphere)} & \multicolumn{2}{c|}{N-body simulation (1/8 sphere)} \\ \hline 
label & redshift range & volume ([Gpc/$h]^3$) & redshift range & volume ([Gpc/$h]^3$) \\ \hline 
$z_0$ & 0-0.629    & 19.9& 0-0.378    & 0.58 \\
$z_1$ & 0.629-0.839 &19.9& 0.378-0.49 & 0.58 \\
$z_2$ & 0.839-1.003&19.9& 0.49-0.575& 0.58\\
$z_3$ & 1.003-1.145&19.9& 0.575-0.646& 0.58\\
$z_4$ & 1.14-1.276&19.9&0.645-0.707 & 0.58\\
$z_5$ & 1.27-1.398&19.9&0.707-0.763 & 0.58\\
$z_6$ & 1.398-1.514&19.9&0.763-0.814 & 0.58\\
$z_7$ & 1.514-1.627&19.9&0.814-0.862 & 0.58\\
$z_8$ & 1.627-1.736&19.9&0.862-0.908 & 0.58\\
$z_9$ & - & - &0.908-0.95 & 0.58\\\hline
\end{tabular}
\label{table:redshift}
\end{center}
\end{table*}

\section{Result}
\label{sec:Result}
In this section, we show the result of our 
KS test and D'Agostino's test 
We estimate linear growth factor parameters 
for different eight redshift bins in Random Gaussian simulation and nine redshift bins in N-body simulation. In pipeline 2, we discard first redshift bin $z_0$. because the non-linear effect is still dominant. 
\subsection{Random Gaussian Result}
\label{ssec:Random Gaussian Result}
\begin{figure}
 \centering
 \includegraphics[width=8cm,clip]{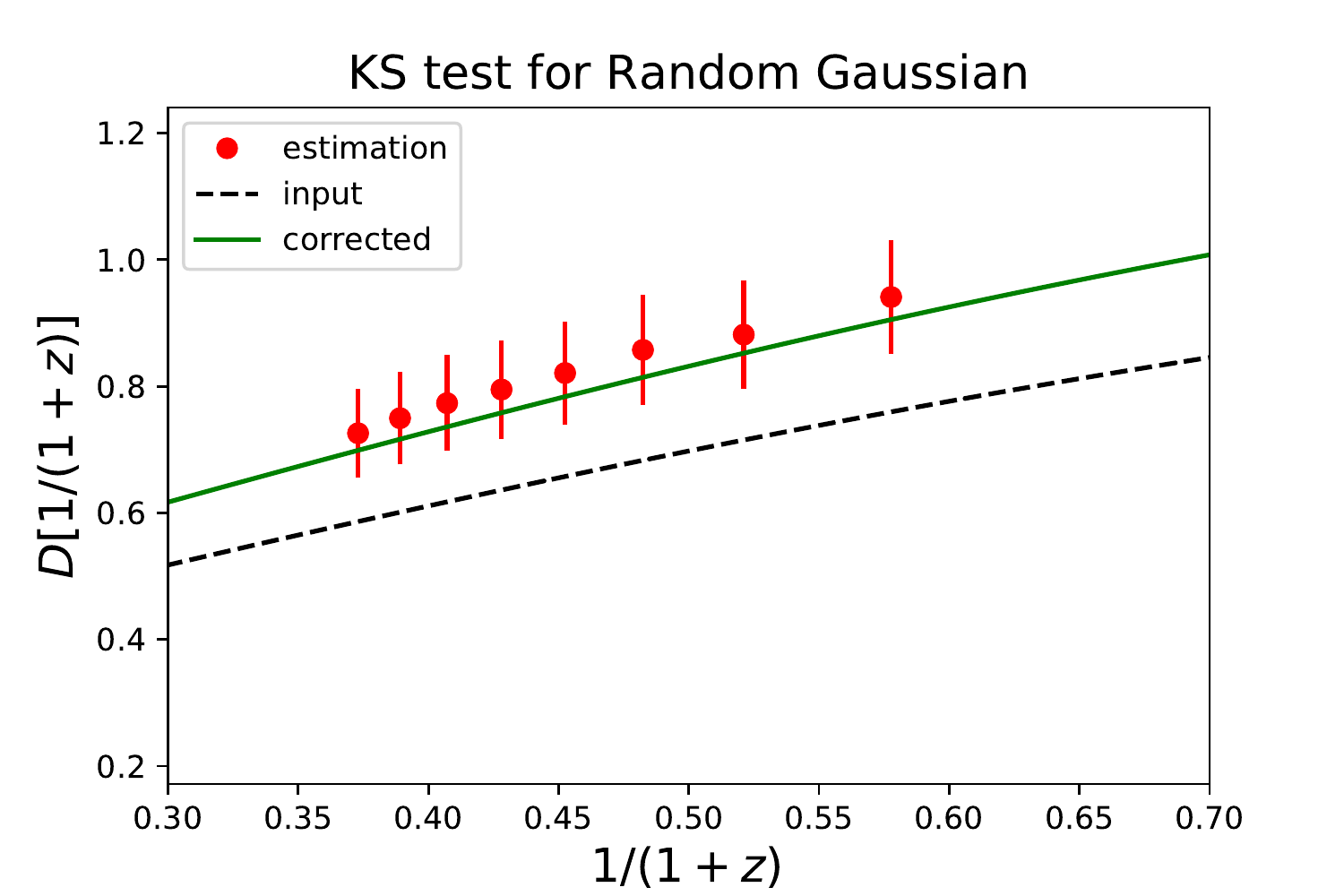}
 \includegraphics[width=8cm,clip]{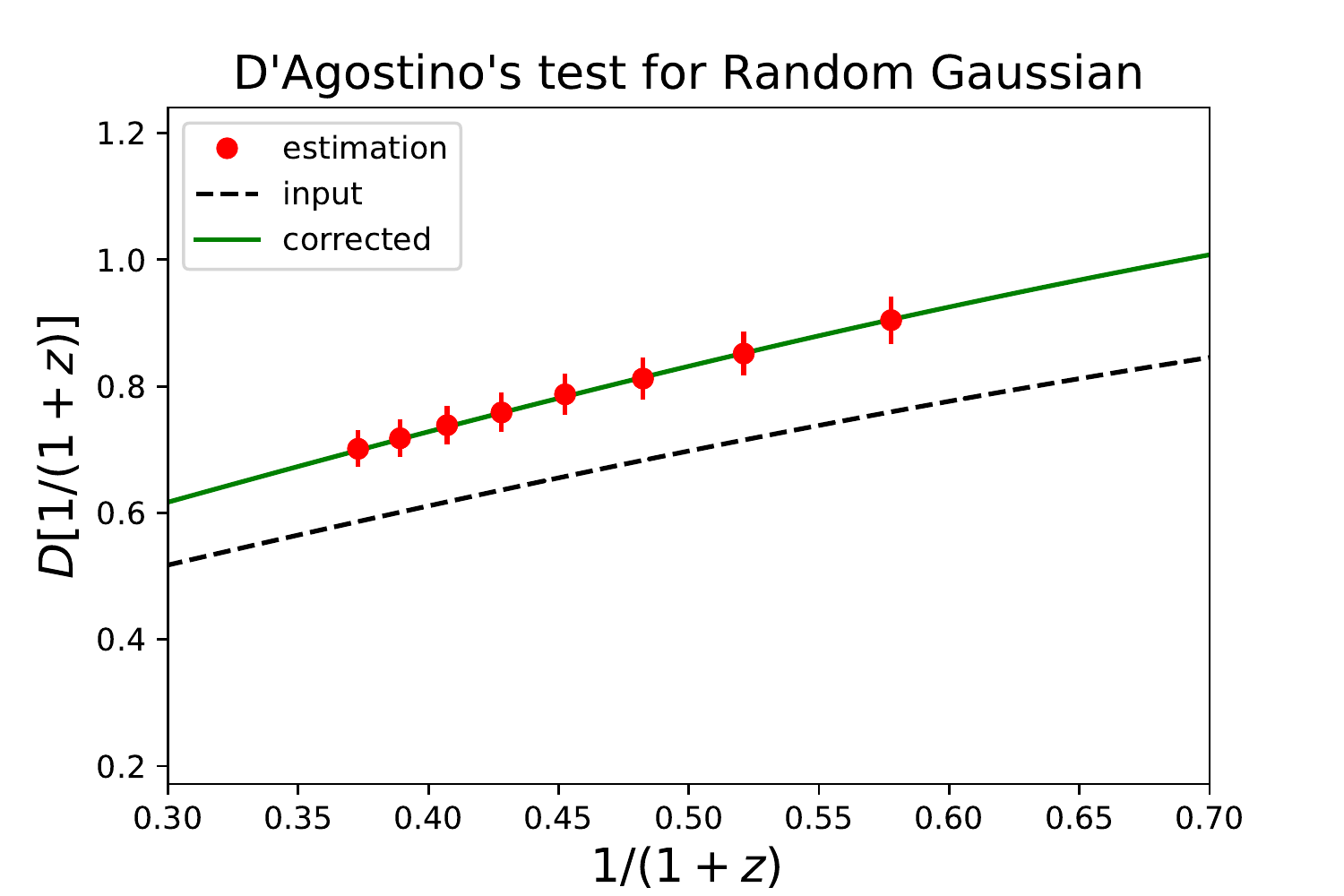}
 \caption{Estimated linear growth factor from random Gaussian simulation as a function of redshift (Red symbols). Upper panel shows the result from KS test and lower panel shows the result from D'Agostino's test. The error bars are computed by calculating variance of MCMC sample data set. 
 Also shown with black dashed line is input growth function assuming $\Omega_{\rm m0}=0.3$ and green line denotes the corrected input model with equation \eqref{correction}.
 }

 \label{MCMC_h_fix_10_R}
\end{figure}

Here we present the results for the Random Gaussian simulation.
In Figure \ref{MCMC_h_fix_10_R}, we show the growth factor $\bm{D}_{\rm est}=[D_{\rm est}(z_1), D_{\rm est}(z_2), \cdots, D_{\rm est}(z_8)]$ estimated using 
KS test (upper panel) and D'Agostino's test (lower panel) 
together with the input growth function.
We see that the estimated growth factors are systematically higher than the input growth function. This is simply due to the fact that we normalise the growth factor to unity at the first bin which has a finite width. For the fair comparison, we correct for the overall amplitude, which can be calculated as
\begin{equation}
    C_{\rm corr}=\int_{z_0}^{z_1} \chi^2(z) D_{\rm true}(z) \frac{dz}{H(z)}.
    \label{correction}
\end{equation}
Therefore, what we have to compare with the estimated values is $\tilde{D}\equiv D_{\rm true}/C_{\rm corr}$. Figure  \ref{MCMC_h_fix_10_R} also shows the corrected input model with green 
line, which perfectly agrees with the estimated values within the error bars.
After estimating parameter $\bm{D}_{\rm est}$, we reconstruct density fluctuation $\delta_{\rm rec}$ using $\bm{D}_{\rm est}$ by simply dividing the observed density fluctuation by the growth factor, 
\begin{equation}
    \delta_{\rm rec}(\bm{x}, z=0)=\frac{\delta_{\rm obs}(\bm{x}, z=z_i)}{D_{\rm est}(z_i)}.
    \label{delta_recon}
\end{equation}
We can quantify how well the reconstruction has been done by the use of KS test. We see that the KS test and D'Agostino's test for the observed $\delta_{\rm obs}$ shows $p$-value $2.76\times10^{-7}$ and $1\times 10^{-117}$ which are sufficiently small to reject the Gaussian distribution, on the other hand, the reconstructed $\delta_{\rm rec}$ shows the $p$-value 0.96 and 0.143 which indicate that our 
method can recover the true Gaussian distribution.  
From Figure \ref{MCMC_h_fix_10_R}, we find that D'Agostino's test provide much better fit for corrected input growth factor than the KS test and the errors are also smaller. 
Therefore, in the next subsection, we will only see the D'Agostino's test for the second (smoothing) approach when we apply the method to the N-body simulation data.

\subsection{N-body Result}
\label{ssec:N-body Result}
%
\begin{figure}
 \centering
 \includegraphics[width=8cm,clip]{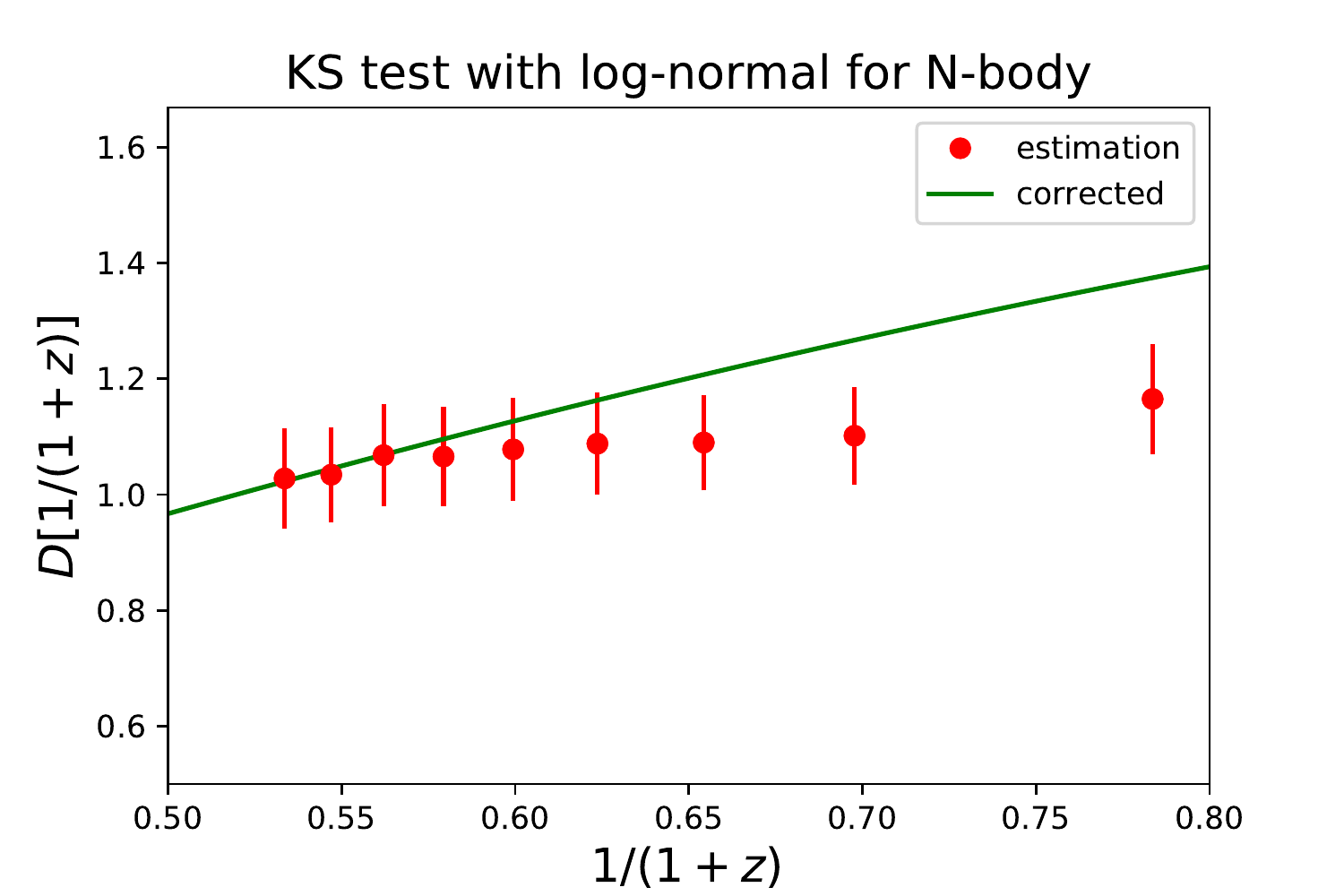}
 \includegraphics[width=8cm,clip]{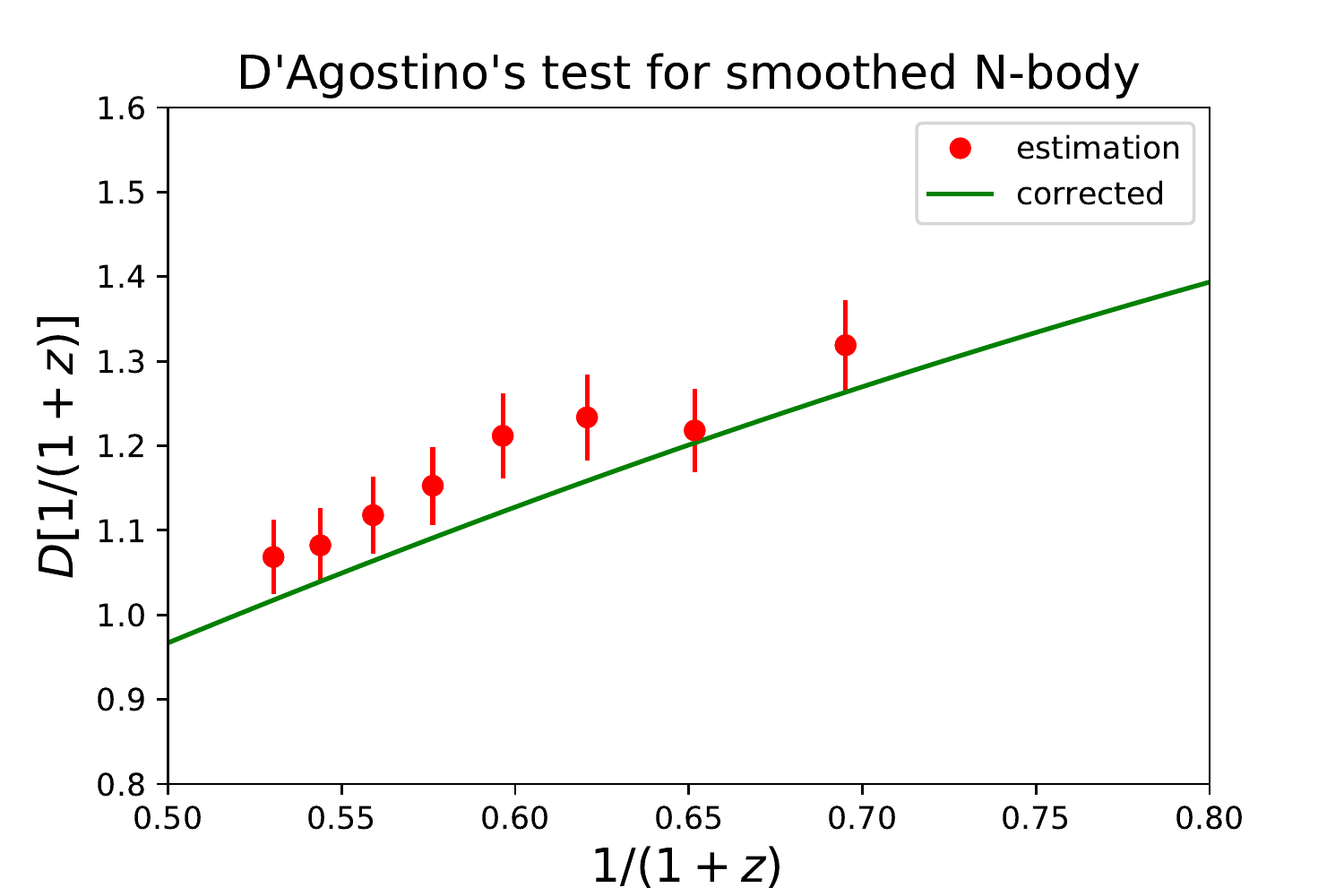}
 \caption{Estimates of the linear growth factor from N-body simulation using two pipelines. Upper panel shows the result from MCMC-KS test using pipeline 1. Lower panel is for pipeline 2 using D'Agostino's test. The data points and color lines are same as in Figure \ref{MCMC_h_fix_10_R}. We calculated $D_{\rm correct}$ using the final redshift bin $z_9$ because we fixed $D(z_9)=1$.
 }
 \label{MCMC_h_fix_10_N}
\end{figure}
In this section, we show the results for the N-body simulation.
As we 
described in section \ref{ssec:Apply for N-body simulation}, we propose two different approaches.
We show the two results corresponding to each method in Figure \ref{MCMC_h_fix_10_N}. Again, the input model is renormalized by the correction factor given in equation \eqref{correction} but unlike the random Gaussian case, we normalise the growth factor at the last redshift bin, $z_9$ and thus the correction factor is calculated within the $z_9$ redshift bin. 
The agreement of the estimated values and input model is worse compared to the random Gaussian case. 
The reason for this disagreement can partly be explained as follows.
In the case of KS test with log-normal distribution, the data may include the non-linear gravitational evolution. It is easily see from the higher order perturbation theory that $\delta(t) = D(t)\delta_{\rm L} + D^2(t) \delta_{\rm L}^2/2 + \cdots$, where $\delta_{\rm L}$ is the initial linear density field, and thus the density fluctuations do not evolve linearly with growth factor. Conversely, our background model is still linear (see Eq. \eqref{delta_recon}). This discordant of the model may cause a notable disagreement of the estimated parameters.
This non-linearity can be mitigated to some extent for the smoothed density field; however, the non-linearity has not been perfectly removed out. Also as seen in Fig. \ref{HIst_comp_N_sm}, the distribution is not well fitted with the Gaussian distribution which may induce the worse constraint compared to the random Gaussian case. It is also notable that the estimated values are systematically higher than prediction. This is partly due to the specific realisation of the N-body simulation. We repeat the same analysis on the 10 different realisations and find that the estimated values over 10 simulation is consistent with the prediction within the 1-$\sigma$ level.

\section{Discussion}
\label{sec:Discussion}
In this section, we 
further explore the analysis based on our results for better understanding of the results.
In section \ref{ssec:covariance}, we will see
the reason why our estimation in N-body simulation is biased above true growth factor (see Figure \ref{MCMC_h_fix_10_N}).
In section \ref{ssec:cosmologyparameter}, we project our model independent constraints on the standard cosmological parameters.

\subsection{Bias of estimation result}
\label{ssec:covariance}
We have observed that the estimated growth factor is more or less biased compared to the input growth function in the N-body simulation. 
As we have already discussed in section \ref{ssec:N-body Result}, this is partly due to the specific realisation of the N-body simulation. Here we quantify the correlation between different redshifts because it highly affects the systematic bias, i.e. if the fluctuation is significantly large at particular redshift bin, the fluctuations around that bin are also enhanced due to the strong correlation among the different bins, if it exists. 

In order to see the correlation among bins, we calculate the correlation coefficient, 
\begin{equation}
    \rho_{ij}\equiv \frac{C_{ij}}{\sqrt{C_{ii}C_{jj}}},
    \label{normalize C}
\end{equation}
where $C_{ij}$ denotes the 
covariance matrix calculated from MCMC samples as 
\begin{equation}
    \label{eq:covariance}
    C_{ij} =
    \frac{1}{N_{\rm samp}} \sum_{k=1}^{N_{\rm samp}}
    \left(D^{(k)}_i - D^{\rm est}_i\right)
    \left(D^{(k)}_j - D^{\rm est}_j\right).
\end{equation}

\begin{figure}
 \centering
 \includegraphics[width=8cm,clip]{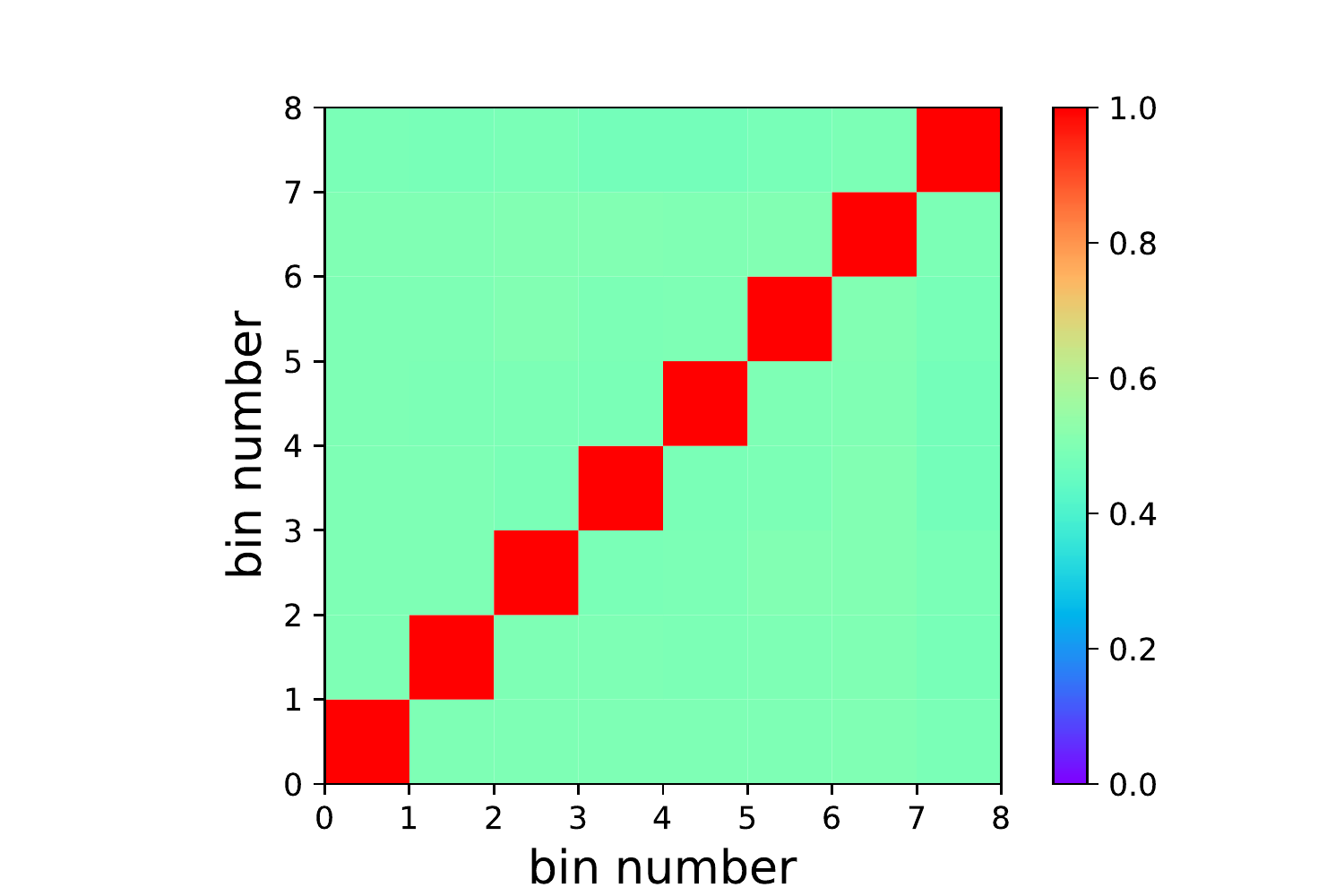}
 \caption{Covariance matrix in N-body simulation. Bin number corresponds to redshift bin number ($z_1,....z_8$). We normalized diagonal components to be 1. The color means that red is 1 and the value becomes smaller when the color goes to be purple. 
 }
 \label{Covariance plot}
\end{figure}

Figure \ref{Covariance plot} shows the correlation 
coefficient.
We see that the cross covariance is fairly large even if two bins are largely separated.
This is because that our method does not rely on the clustering of the density field but solely on the fluctuation amplitude and its distribution at every grid point. Suppose that the amplitude of at given redshift bin is slightly higher, amplitudes for other bins also need to get higher in order for the entire sample to be a Gaussian distribution. Furthermore, the effect of amplification of the fluctuations is same for neighbouring bins or farther distant bins. That makes our cross covariance quite uniform across different redshift bins.

As our covariance matrix is positive definite symmetric matrix, it can be always diagonalised as
$\bm{\Lambda}=\bm{U}^{-1}\bm{C}\bm{U}$, 
where $\bm{\Lambda}$ is  a diagonal matrix which consists of eigenvalues of $\bm{C}$ and $\bm{U}$ is an orthogonal matrix which consists of eigenvalues of $\bm{C}$.

\begin{figure}
 \centering
 \includegraphics[width=8cm,clip]{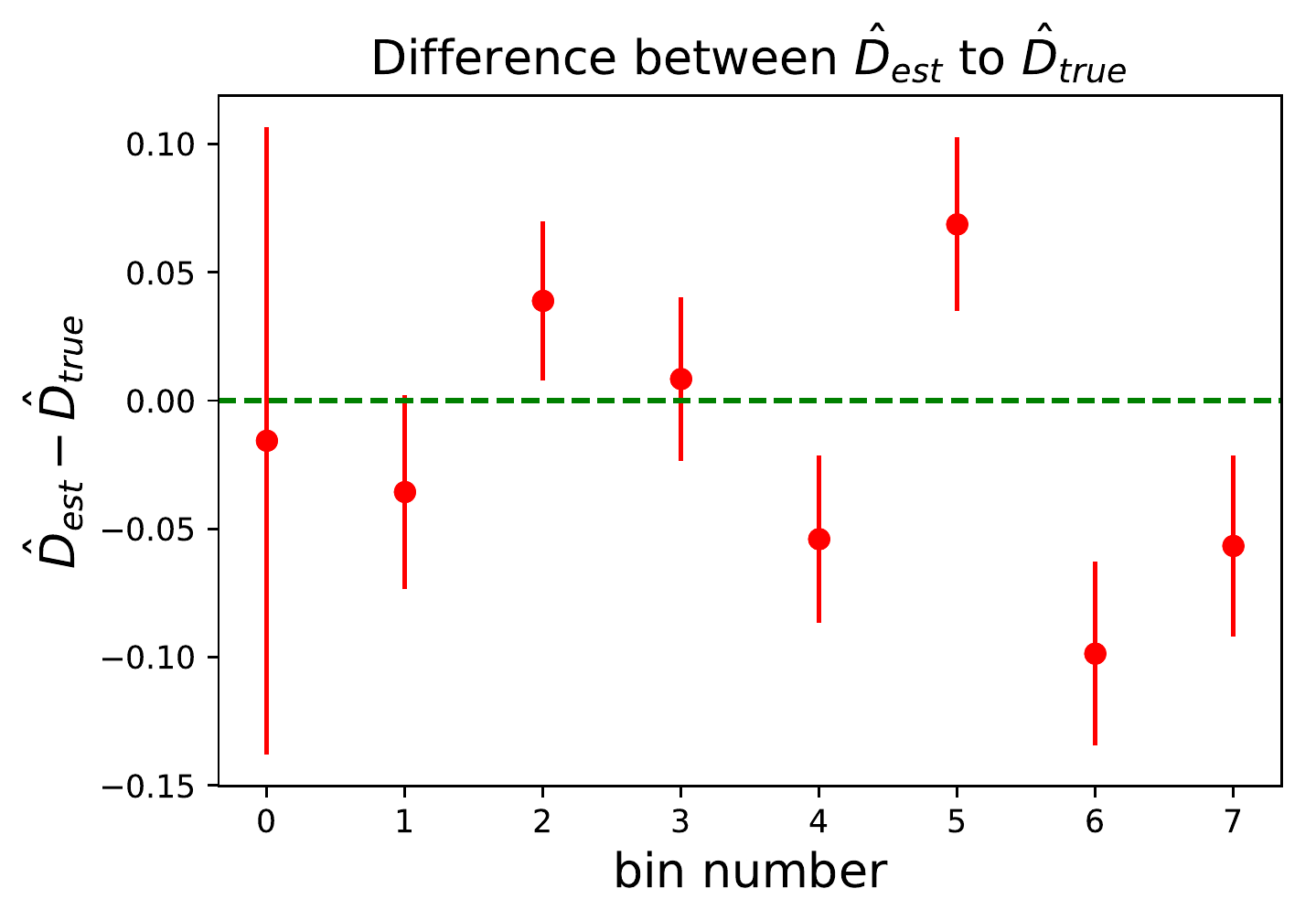}
 \caption{Difference between the true values and the estimated values in independent space using N-body simulation. Green dashed line shows where the vertical value is 0. Red dots are calculated  from equation \eqref{D_independent} with the $1\sigma$ variance which is a square root of the eigenvalue sorted in descending order.}
 
 \label{fig:independent_params}
\end{figure}

Now our estimated parameters are linearly transformed as
\begin{equation}
    \left\{
        \begin{alignedat}{4}
        \hat{\bm{D}}_{\rm est} &=\bm{U}\bm{D}_{\rm est}\\
        \hat{\bm{D}}_{\rm true} &=\bm{U}\tilde{\bm{D}}=\bm{U}\frac{\bm{D}_{\rm true}}{C_{\rm corr}},
        \end{alignedat}
    \right.
    \label{D_independent}
\end{equation}
Figure \ref{fig:independent_params} shows the difference between decorrelated estimated parameter and input.

We now see that the systematic bias seen in Figure \ref{MCMC_h_fix_10_N} has been totally gone after decorrelated parameters and the $\hat{D}_{\rm est}$ is randomly scattered around $\hat{D}_{\rm true}$.

Another possibility to generate the bias is 
smoothing. 
We smooth the data in Fourier space in the cubic simulation box;however, we only have the meaningful data within the sphere centered at the bottom corner of the cube. Since we set the fluctuations outside the sphere zero, the smoothing reduces the amplitude of fluctuations near the boundary.
In N-body simulation, we fix the growth factor at the final redshift bin $D(z_9)=1$ corresponding to the boundary and therefore the estimated growth factors at the lower redshift bins are higher than we expected. 
We see that this effect depends on the realization of the fluctuation and the smoothing does not fully account for the systematic bias.
We conclude that the systematic bias is due to the strong correlation among the parameters in different bins and boundary effect of smoothing.

\subsection{Cosmological Parameter}
\label{ssec:cosmologyparameter}
Apart from the amplitude of the growth factor, the redshift evolution of growth factor is more informative about the cosmological model \citep[e.g.][]{2008PhRvD..77h3508D, 2013PhRvD..88h4032X,2010JCAP...04..022D}. Now we assume that the structure formation is totally described by the linear perturbation theory within a regime of general relativity. Without spatial curvature, the growth function can be parameterized by the single parameter, $\Omega_{\rm m}$.
In the case of absence of cosmological constant (i.e. $\Omega_{\rm m}=1$), the linear growth function can be expressed as,
\begin{equation}
    D(z)\propto \frac{1}{1+z}=a,
    \label{de Sitter solution}
\end{equation}
Compared with the Einstein-de Sitter model, $\Lambda$CDM model includes the influence of accelerating expansion. This suppresses the growth of density fluctuation.\par

\begin{figure}
 \centering
 \includegraphics[width=8cm,clip]{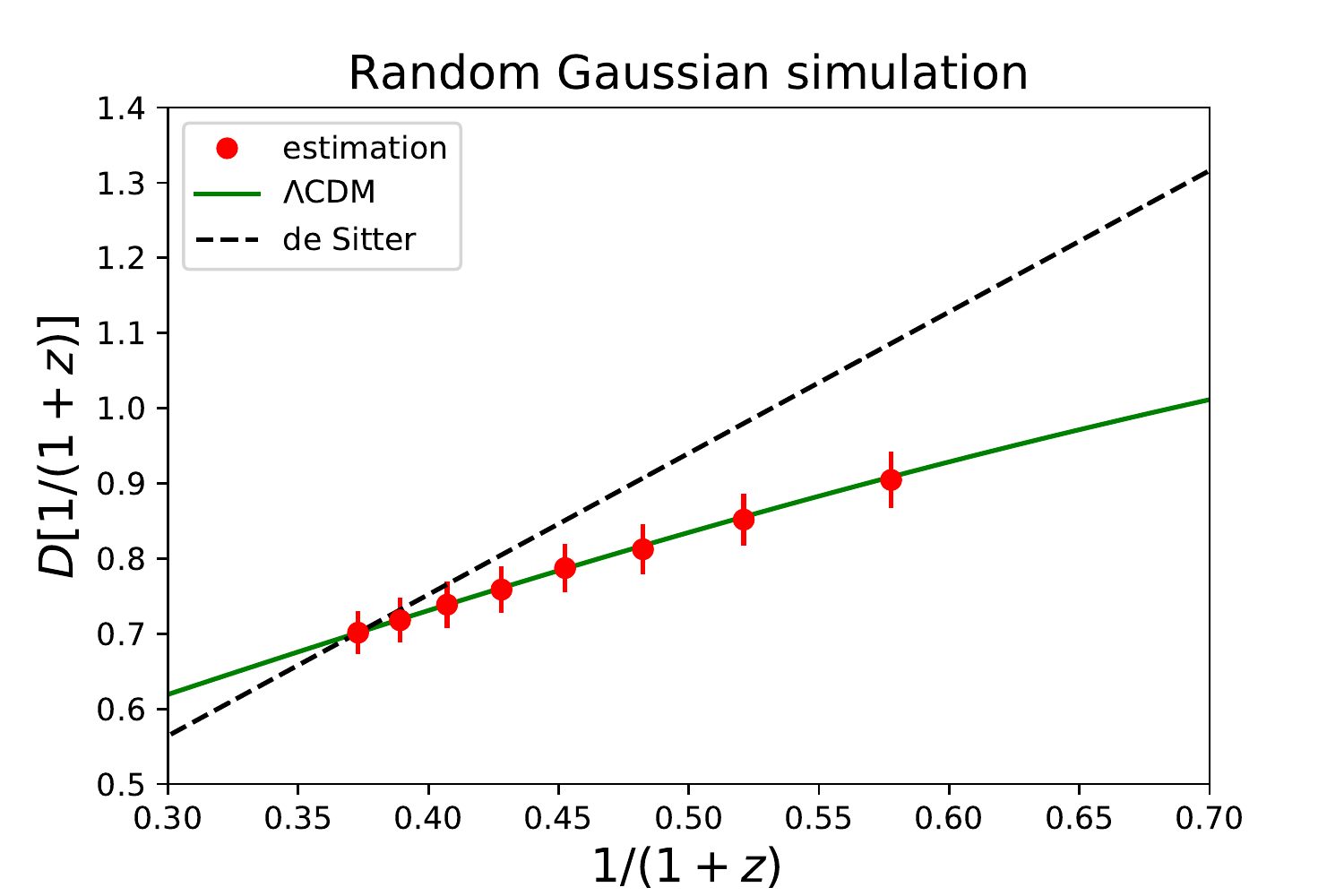}
 \includegraphics[width=8cm,clip]{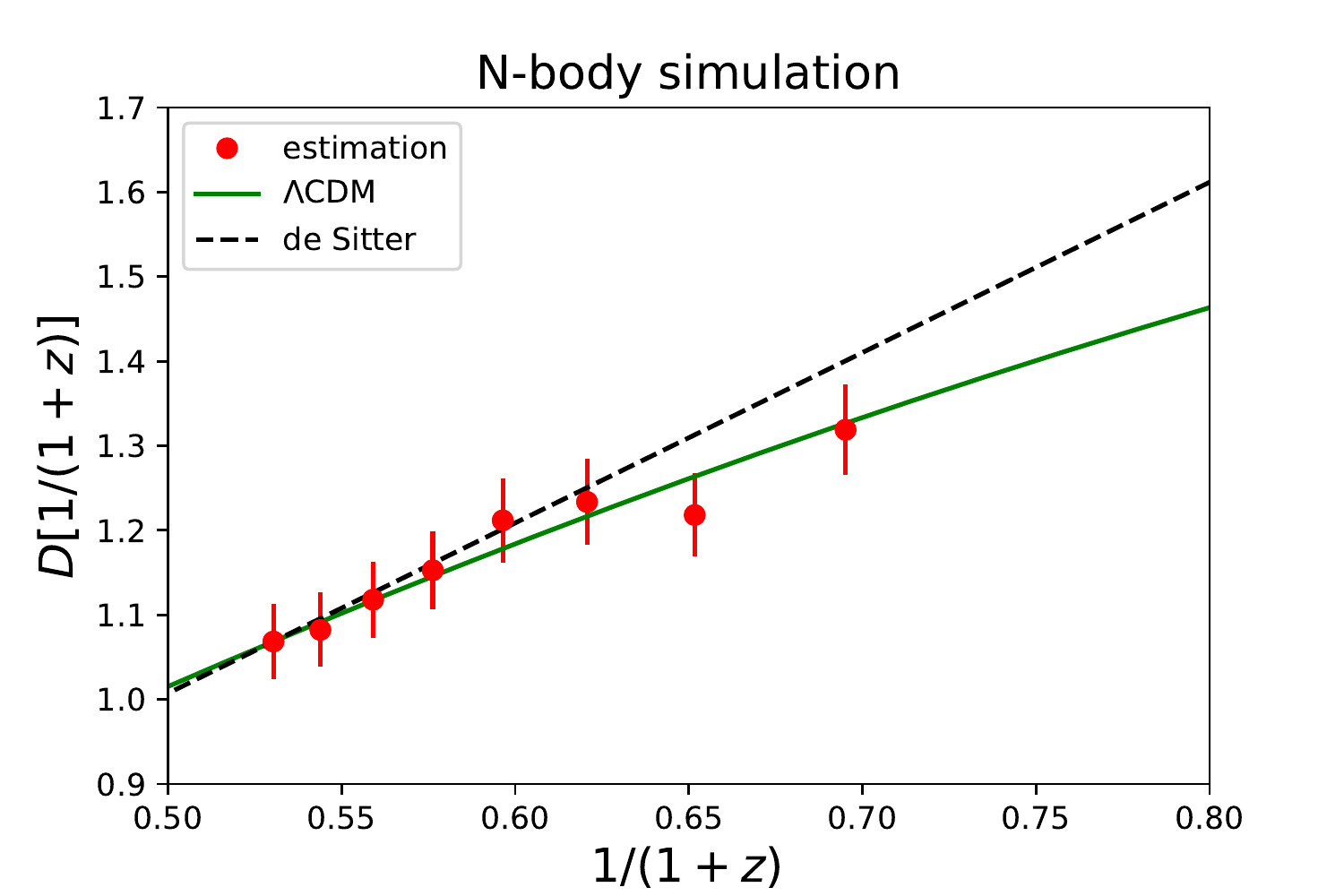}
 \caption{Comparison with the $\Lambda$CDM model and Einstein de-Sitter model using D'Agostino's test. Green line shows $D_{\rm true}$ rescaled to fit $D_{\rm est}$ of the final redshift bin assuming $\Lambda$ CDM model and black dashed line shows the growth factor of Einstein-de Sitter model. Upper and lower panel show Random Gaussian simulation, smoothed N-body simulation respectively.}
 \label{Dark energy R}
\end{figure}

Now, suppose we obtain the linear growth factor $\bm{D}_{\rm est}$ using our method and get the same result as in section \ref{sec:Result}. With our non-parametric method, we can study whether the component of our universe is only matter or not without assuming any prior.
In Figure.\ref{Dark energy R}, we compare two models with our estimation results using Random Gaussian and N-body simulations. As you can see, the estimated points do not follow the de-Sitter model. For example in Random Gaussian simulation, dashed line is beyond $1\sigma$ errors of some data points and also in the N-body simulation, dashed line does not provide a good fit at the two lowest redshift bins. We can see that the growth of the density fluctuation is suppressed compared with the de-Sitter model. It will constitute evidence of dark energy without assuming any cosmological model.  

Next, if we restrict our interest to the $\Lambda$CDM model, then we can estimate cosmological parameters using the result from our non-parametric method. Linear growth factor depends on $\Omega_{\rm m0}$. Now, we show how we can constrain $\Omega_{\rm m0}$ using Random Gaussian simulation result from section \ref{ssec:Random Gaussian Result}. To compare our result, we define the growth factor model $D_p$ parameterized by $\Omega_{\rm m0}$ and the overall amplitude $A_D$ of the 
growth factor as
\begin{equation}
    D_p(\Omega_{\rm m0}, A_D, z)=A_DD_{\rm \Lambda CDM} (\Omega_{\rm m0}, z).
    \label{parametrize D}
\end{equation}
Then, we can calculate $\chi^2(\Omega_{\rm m0}, A_D)$ using covariance matrix 
\begin{equation}
    \chi^2(\Omega_{\rm m0}, A_D)
    =
    (\bm{D}_{\rm est}-\bm{D}_p)^T
    \bm{C}^{-1}
    (\bm{D}_{\rm est}-\bm{D}_p),
    \label{chi_square}
\end{equation}
where the covariance matrix is given by equation \eqref{eq:covariance}.
Then we can derive the posterior distribution of $\Omega_{\rm m0}$ by marginalizing over $A_D$,

\begin{equation}
    P(\Omega_{\rm m0})\propto\int dA_D\ {\rm exp}\left[-\frac{\chi^2(\Omega_{\rm m0}, A_D)}{2}\right].
    \label{p_omegam0}
\end{equation}
\begin{figure}
 \centering
 \includegraphics[width=8cm,clip]{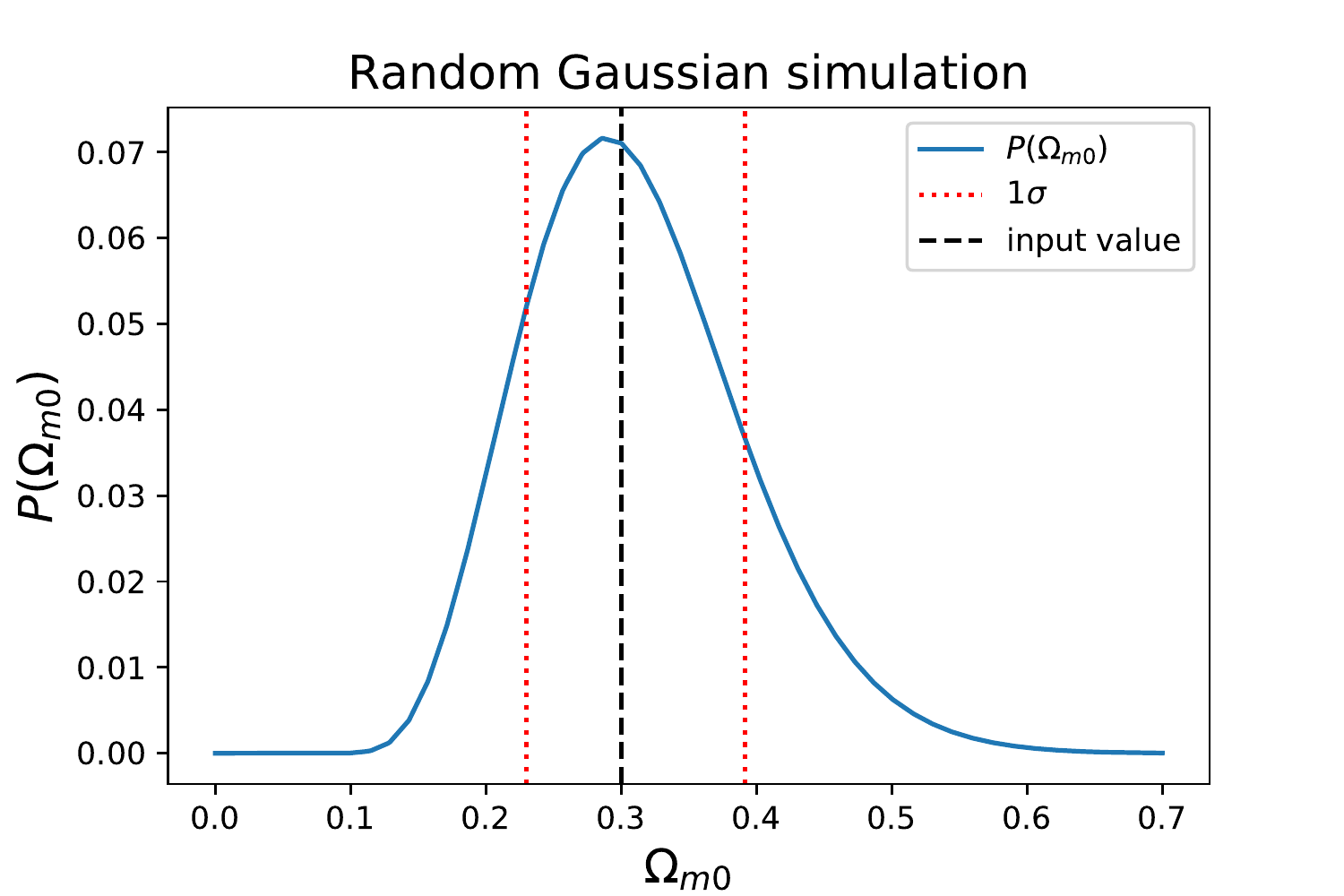}
 \includegraphics[width=8cm,clip]{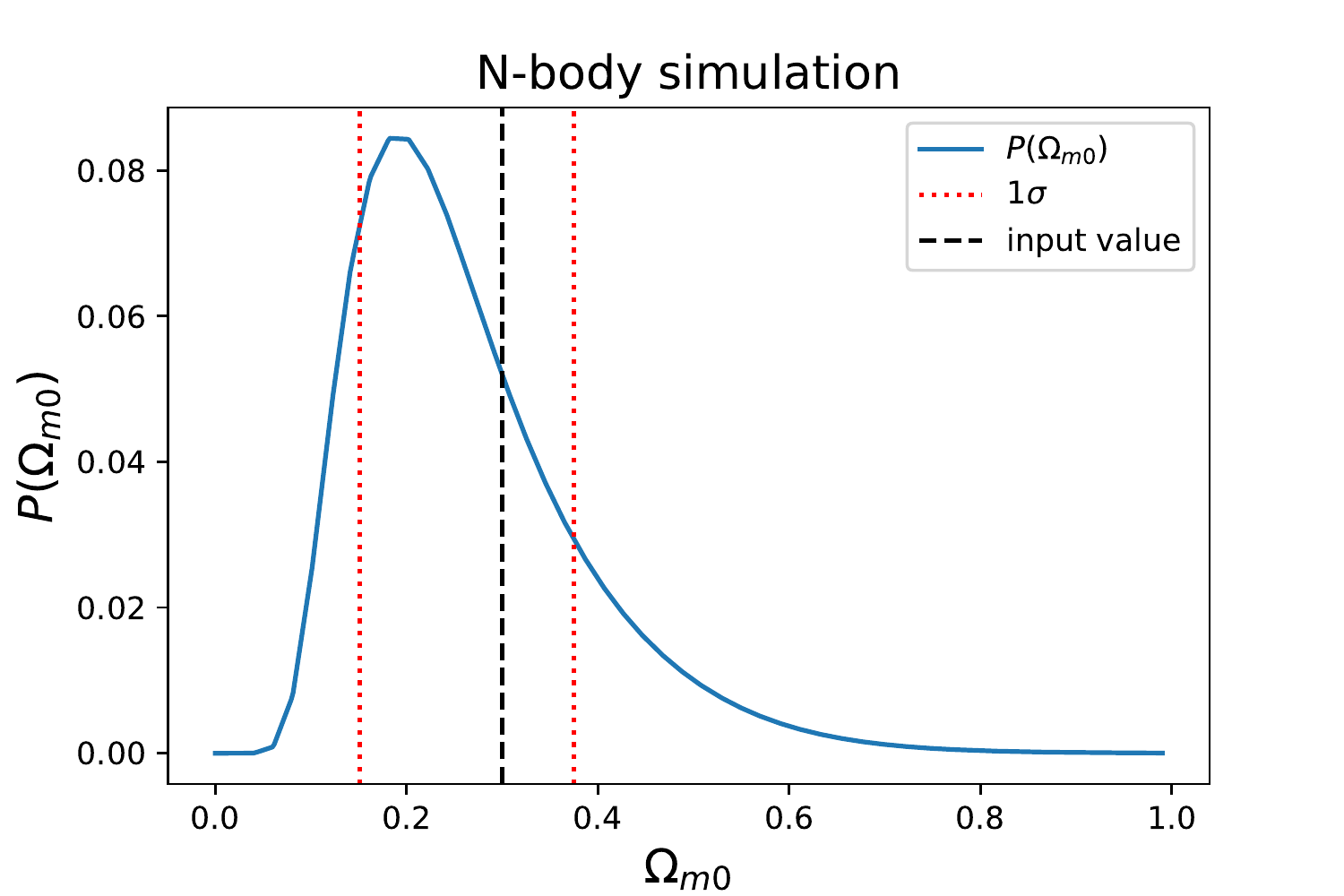}
 \caption{
 Posterior distribution function of $\Omega_{\rm m0}$ after marginzlized over $A_D$. Upper panel shows the result from Random Gaussian simulation and lower panel shows the result from N-body simulation. Red dotted lines show the upper and lower $1\sigma$ limits. Black dashed line denotes the input value of the simulation, $\Omega_{\rm m0}=0.3$ in Random Gaussian simulation and $\Omega_{\rm m0}=0.311$ in N-body simulation.
 }
 
 \label{Omega_m}
\end{figure}
We show the result in Figure \ref{Omega_m}. We obtain parameter value $\Omega_{\rm m0}=0.2862^{+0.1038}_{-0.0562}$ ($68\%$C.L.) from Random Gaussian simulation and $\Omega_{\rm m0}=0.1826^{+0.1924}_{-0.0318}$ ($68\%$C.L.) from N-body simulation (pipeline 2). 
\begin{table}
  \begin{center}
    \caption{Parameter estimation result. $\Omega_{\rm m0}$ calculated from equation \eqref{p_omegam0} marginalized around $A_D$. Lower and Upper limit correspond to $1\sigma$ limits. Input value is $\Omega_{\rm m0}=0.3$ in Random Gaussian simulation and $\Omega_{\rm m0}=0.311$ in N-body simulation.}
    \label{estimation table}
    \begin{tabular}{|c|c|c||c|} \hline
      simulation&lower limit & mean & upper limit \\ \hline \hline
      Random Gaussian& 0.230 & 0.2862 & 0.399\\
      N-body& 0.151 & 0.183 & 0.375\\\hline
    \end{tabular}
  \end{center}
\end{table}
As we can see, we estimate the input value within $1\sigma$ error.\par
Next, we will demonstrate that our result can be used to constrain the linear growth rate $f$ defined as the time derivative of the linear growth factor
\begin{equation}
    f(a)=\frac{d\ln D}{d\ln a}.
    \label{e:growth rate}
\end{equation}
It can be calculated by numerical integration, but approximately it is often expressed as
\begin{equation}
    f(a)=\Omega^\gamma_{\rm m}(a),
    \label{e:de}
\end{equation}
where $\gamma$ is so called growth rate index parameter. This parameter depends on gravitational theories. In the usual growth history (within the general relativity framework), the parameter takes the value $\gamma\sim6/11$. Any deviation from this value would imply the need for modified gravity. To estimate the $\gamma$ parameter, we parametlize $\Omega_{\rm m0}$, $\gamma$ and the amplitude of linear growth factor $A_D$ as
\begin{equation}
    D_{\rm model}(a)=A_D{\rm exp}\left(-\int^1_a\frac{f(a, \Omega_{\rm m0}, \gamma)}{a'}da'\right).
    \label{e: integrate D}
\end{equation}
We will apply this $D_{\rm model}(a)$ to our estimated result $\bm{D}_{\rm est}$ from N-body simulation pipeline 2, and calculate $\chi^2(\Omega_{\rm m0}, \gamma, A_D)$ in a similar way as in the equation \eqref{chi_square}. Then we can derive the joint posterior distribution of ($\Omega_{\rm m0}, \gamma$) by marginalising over $A_D$.\par
\begin{figure}
 \centering
 \includegraphics[width=8cm,clip]{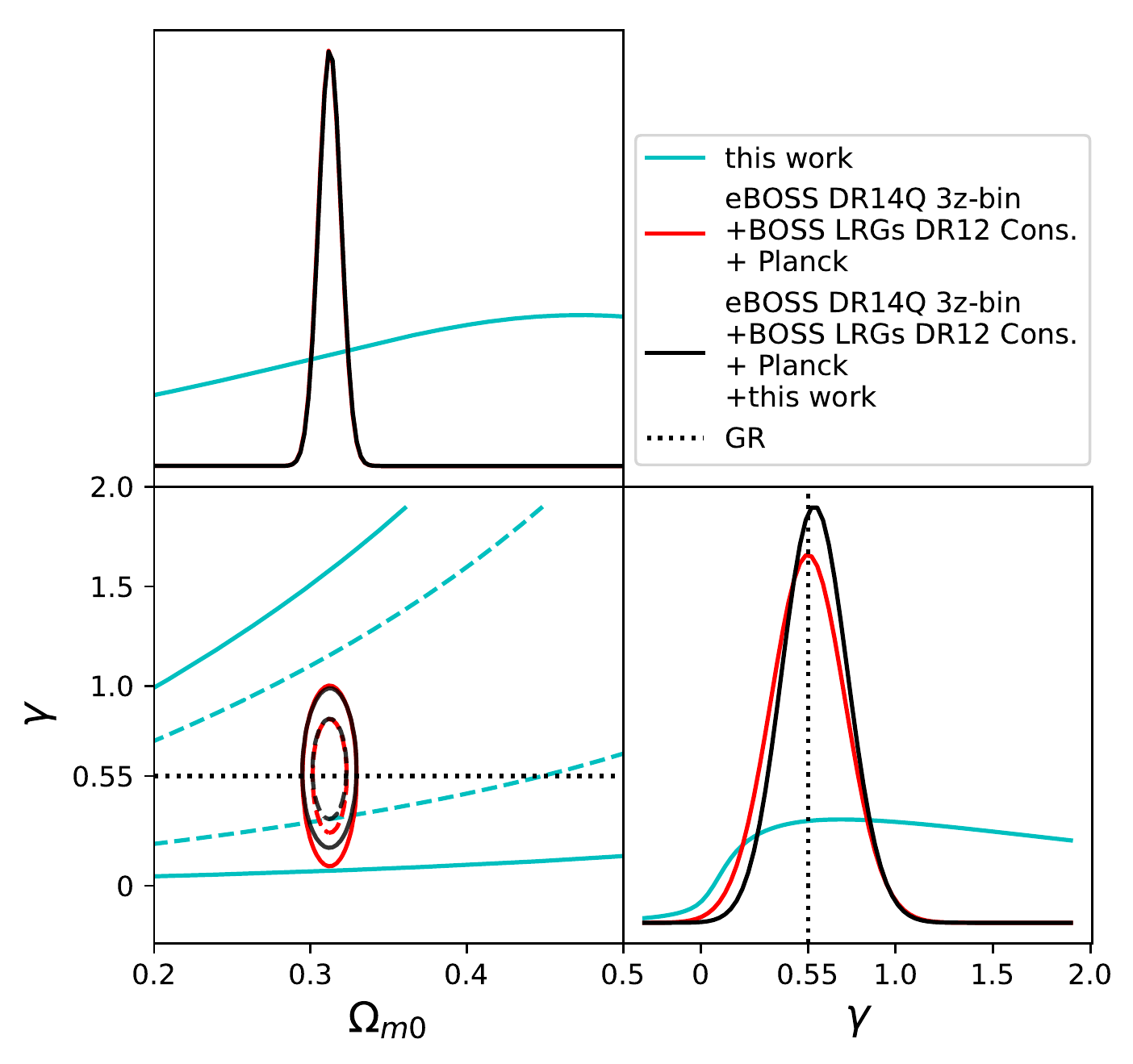}
 \caption{Constraints on $\Omega_{\rm m0}$ and $\gamma$ from our method. Lower left panel shows the $1$ and $2$ sigma contours of the joint probability of ($\Omega_{\rm m0}, \gamma$) and lower right panel shows the probability of $\gamma$ after marginalizing over $A_D$ and  $\Omega_{\rm m0}$. Upper panel shows the probability of $\Omega_{\rm m0}$ after marginalizing over $A_D$ and $\gamma$. The cyan lines are the result using uniform prior for all parameters and red line is the result from eBOSS DR14Q+BOSS DR12 LRGs+Planck \citep{2018MNRAS.477.1604G}. Black line is the result of our estimation using  eBOSS DR14Q+BOSS DR12 LRGs+Planck result as the prior. In the contour plot, the solid and dashed lines correspond to $2\sigma$ and $1\sigma$ errors, respectively. The dotted lines are the predicted $\gamma$ from the Einstein gravity.}
 \label{f:gamma_const}
\end{figure}

We show the result in Figure.\ref{f:gamma_const}. When we estimate the joint probability without using any priors on the parameters, we can not estimate both the parameters $\Omega_{\rm m0}$ and $\gamma$ because they degenerate with each other. To see the impact of our method, we use as the prior for the joint probability of $(\Omega_{\rm m0}, \gamma)$ the eBOSS DR14Q, BOSS DR12 LRGs and Planck results \citep{2018MNRAS.477.1604G, 2017MNRAS.470.2617A, 2016A&A...594A..13P}. Then, we obtain a constraint about $\gamma$. As we can see from Figure.\ref{f:gamma_const}, the constraint becomes a little stronger after adding our result. Actually, \cite{2018MNRAS.477.1604G} obtains the constraint as $\gamma=0.55^{+0.19}_{-0.19}$ ($68\%$C.L.) and after adding our result, we obtain $\gamma=0.58^{+0.17}_{-0.16}$ ($68\%$C.L.). From the above result, we can say that our new method will contribute to the study of gravity theories. 

\subsection{Error budget}
Here we have a brief discussion about the error 
budget.
In our analysis, both the cosmic variance and shot noise may affect the errors. In order to discriminate the error contents, we perform two different analyses. First, we repeat our analysis with reducing the volume by half keeping the same number density of the N-body particles. If we consider a galaxy survey,
this is equivalent to reducing the observation volume by half with a fixed number density of galaxies $n_{\rm g}\sim 0.014\ h^3{\rm Mpc}^{-3}$. 
This number density is comparable to that expected for the "Euclid wide survey" \citep{2018LRR....21....2A}, which will observe 15,000 ${\rm deg}^2$ of the sky for the redshift range of $0.7<z<1.8$.
In this case, we find that the errors in the estimates of $D$ 
become
larger by a factor of $\sim \sqrt{2}$, which is a simple scaling 
of the cosmic variance limited regime.
Second, we analyse the data with the same volume, but reducing the number of N-body particle by half. In this case, interestingly, we find that the errors in the estimates of $D$ get smaller by a factor of $0.95$. 
This might be explained as follows: 
omitting the high-$k$ modes by reducing the number of particles makes the distribution of density perturbation 
closer to the Gaussian, and thus the 
the data is more eligible for the D'Agostino's test.
This means that there might be 
the most appropriate scale
for the spatial resolution of density perturbations 
to perform the
D' Agostino's K-squared test. In fact, if we set $k_{\rm cut}$ larger than $0.07\ h/{\rm Mpc}$, the growth factors estimated by our tests get biased with larger error bars because of the nonlinearity on small scales which skews the distribution from Gaussian one. On the other hand, if $k_{\rm cut}$ is set smaller than $0.07\ h/{\rm Mpc}$, the error bars becomes larger as well because the number of the available $k$ mode decreases. Our simple investigation shows that the scale of the sweet spot is around $k\sim 0.07$ $h$/Mpc, but 
finding
that scale precisely 
might be done prior to the practical measurement by use of mock simulations and 
is beyond the scope of this paper. Finally, we note that in our fiducial analysis we use $1/8$ of the full sky; and therefore we may reduce the size of errors in principle by a factor of $1/\sqrt{8}$ in an ideal full sky analysis.

\section{Summary}
\label{sec:Conclusion}
In this paper, we proposed a novel method to constrain the growth factors for a wide range of redshifts and to reconstruct the density field at present epoch simultaneously, by using the one point distribution function of the density field. Even if the initial density field obeys a Gaussian distribution, the observed density field on our light cone is not a Gaussian one since density fluctuations at different redshifts obey Gaussian distributions with different variances. We utilise the Kolmogorov-Smirnov test and D'Agostino's K-squared test to estimate the growth factors based on the fact that the linearly evolved density field normalized by the growth factors should obey a Gaussian distribution with a single variance. This approach is completely independent from the standard analyses based on the isotropic and anisotropic power spectra, and may offer an independent measure of the cosmological model.

In order to verify our method, we first applied KS and D'Agostino's tests to the
random Gaussian field that grows linearly in time. From the test using the random Gaussian data we found that D'Agostino's test is statistically stronger than the KS test.\par
Next, we considered the N-body simulation data to investigate how the non-linear gravitational growth of density field would affect the results of our normality tests.
We considered two different pipelines. The first is to use the KS test setting the log-normal distribution as the target distribution because the non-linearly evolved density field is approximately described by a log-normal distribution. The second is using only large scale modes of the density field to treat its distribution as approximately Gaussian one and applying D'Agostino's test. As a result, we found that we can not estimate "linear" growth factors from the density fluctuation which is evolved "non-linealy" from pipeline 1. And from pipeline 2, we found that we can estimate the time evolution of the linear growth factor but not the overall amplitude. We also found that the reason of the bias of our estimation may be explained by the smoothing effect and correlation between the estimated growth factors.

Finally, we demonstrated how we can constrain cosmological models using our estimation results. Our non-parametric result can be used to show the evidence for the existence of dark energy and to estimate $\Omega_{\rm m0}$ without utilizing the power spectra.  Our simulation setup was comparable to the Euclid wide survey, but we assumed the number density of the galaxy is the same as that of the dark matter particles, and therefore we have to consider more realistic cases (for example, galaxy bias, redshift-space distortion, and so on).  Although many challenges exist for applying it to the actual cosmological problem, the method proposed here provides a new way of testing the growth of large scale structure in a non-parametric way. 

\section*{Acknowledgements}
This work is supported in part by JSPS KAKENHI Grant Numbers
18K03616, 15H05890, 17H01110. We thank Naoshi Sugiyama, Nabila Aghanim, Shiro Ikeda, and Arman Shafieloo for useful discussions.

\bibliographystyle{apsrev}
\bibliography{bibdata2}

\end{document}